\documentclass[twocolumn,showpacs,preprintnumbers,amsmath,amssymb,lettersize]{revtex4}

\usepackage{graphicx}
\usepackage{dcolumn}
\usepackage{bm,bbm,amsmath}

\usepackage{amsthm} 
 
\newcommand{\ket}[1]{\mbox{$ | #1 \rangle $}} 
\newcommand{\bra}[1]{\mbox{$ \langle #1 | $}} 
\newcommand{\braket}[1]{\mbox{$ \langle #1 \rangle $}}

\begin{document}

\title{Testing Quantum Devices: Practical Entanglement Verification in Bipartite Optical Systems} 
 
\author{Hauke H\"aseler, Tobias Moroder, and Norbert L\"{u}tkenhaus} 
\affiliation{Quantum Information Theory Group, Institut f\"{u}r Theoretische Physik I, \& Max-Planck Research Group, Institute of Optics, Information and Photonics, Universit\"{a}t Erlangen-N\"{u}rnberg, Staudtstra{\ss}e 7/B2, 91058 Erlangen, Germany \\}
\affiliation{Institute for Quantum Computing \& Department of Physics and Astronomy, University of Waterloo, University Ave.~W.~N2L 3G1, Canada}
 
\date{\today} 
 
\begin{abstract} 
We present a method to test quantum behavior of quantum information processing devices, such as quantum memories, teleportation devices, channels and quantum key distribution protocols. The test of quantum behavior can be phrased as the verification of effective entanglement. Necessary separability criteria are formulated in terms of a matrix of expectation values in conjunction with the partial transposition map. Our method is designed to reduce the resources for entanglement verification. A particular protocol based on coherent states and homodyne detection is used to illustrate the method. A possible test for the quantum nature of memories using two non-orthogonal signal states arises naturally. Furthermore, closer inspection of the measurement process in terms of the Stokes operators reveals a security threat for quantum key distribution involving phase reference beams.
\end{abstract}

\pacs{03.67.Hk, 42.50.Ex, 03.67.Dd}

\maketitle

\section{INTRODUCTION}

Quantum communication protocols can perform tasks that are not possible classically such as quantum key distribution \cite{bennett84a, ekert91a}, teleportation \cite{bennett93a}, and quantum memory \cite{kozhekin00a}. Additionally, quantum communication can outperform possible classical counterparts, as is the case for superdense coding \cite{bennett92e}.

In experimental realizations (see, e.g., Refs.~\cite{julsgaard04a, bouwmeester97a, stucki02a}), imperfections are inevitable, and the advantage over a classical strategy may be lost. For example, a classical strategy for a quantum memory performs a quantum measurement on the incoming signal, stores the result classically, and prepares a quantum state corresponding to the stored result as output. For quantum key distribution (QKD), the classical strategy to compare against is an intercept-resend attack. A typical test for quantum memories and teleportation devices is to monitor the fidelity of the stored or transmitted quantum states and to compare the observed value to the theoretical limit for classical strategies \cite{braunstein98a, kozhekin00a, hammerer05a}.

 Especially in infinite-dimensional systems, however, knowing which fidelities are classically achievable is non-trivial. The question is answered for Gaussian distributions of coherent input states in Ref.~\cite{hammerer05a}, but an answer for general input states is still outstanding. Also, from an experimental viewpoint, a continuum set of test states, such as a Gaussian distribution, is problematic, and a test involving only a small number of input states would be preferable. 

The test we develop here does not make direct use of any figure of merit and has a simple structure: The quantum device is tested on a set of non-orthogonal signal states, and the corresponding output states are then measured in one of a few measurement settings. In some situations, for example, in testing QKD devices or quantum teleportation setups, there might be two parties, possibly at distant locations, one party (Alice) preparing the input states, and the other (Bob) performing measurements on the output states. In this case, the two parties might make use of classical public communication to collect the necessary information to evaluate the test. In other situations, like testing a quantum memory, the preparation and measurement can be performed by one and the same party. 

The test procedure can be reformulated using an entanglement-based description. In this formulation, an alternative description of the state preparation is used \cite{bennett92c}, where Alice creates a bipartite, entangled state and sends one subsystem to Bob. By performing a suitable measurement on her own subsystem, she can effectively prepare the desired signal state. In general, before Alice's and Bob's measurements, we will have an effective entangled state shared by the two parties. As far as the action of the quantum device and Bob's measurements are concerned, both formalisms are equivalent. 

The main test idea is that classical devices, in the special meaning introduced above, will always destroy any entanglement between Alice and Bob in the entanglement-based description. Conversely, if entanglement in the underlying state can be shown from the measurement data, then the minimal requirement of actual quantum behavior of the device relevant to quantum communication is demonstrated. Note that the notion of classicality tested here and relevant to the use of the devices in a quantum communication context differs from the notion of classicality often discussed in a quantum optical context \cite{mandel86a}.

To use entanglement verification as a first test of a protocol has been proposed in the context of quantum key distribution \cite{curty04a,curty05a}. The underlying idea has also been used to show ultimate limits for specific QKD protocols \cite{curty07a,curty07b}. For other tasks, such as the storage of quantum states, translation to an entanglement-based picture and verification of quantum correlations may provide an easily implementable test.\\

We present in this paper a method to verify entanglement especially suitable for bipartite optical systems, which is in principle not limited to a particular Hilbert space dimension or specific choices of measurements. We define the bipartite expectation value matrix, which allows to detect entanglement even if only partial information on the system is available.

Section \ref{sec:method} gives the general definitions and properties connected to the verification method. Its workings are exemplified in Section \ref{sec:apply} on a discrete-continuous cryptography protocol with homodyne detection. Furthermore, the applicability of the verification method to quantum memories is discussed. Section \ref{sec:evmstokes} revisits the same protocol and takes into account the phase reference necessary for the detection process. The resulting description in terms of the Stokes operators will serve as a second example of the applicability and limitations of the entanglement verification method and raise questions about the security of quantum key distribution based on homodyne detection. We show that the security may be compromised by changes in the phase reference intensity. Section \ref{sec:addm} investigates remedies in the form of additional measurements. Additional results for the presence of channel losses are given in Section \ref{sec:loss}. We close with a conclusion in Section \ref{sec:conclude}.

\section{General Entanglement Verification Method}\label{sec:method}
In this section we address the problem of entanglement verification in the situation, in which the measurements performed on a bipartite system are not tomographically complete, i.e., the density matrix of the underlying state cannot be completely characterized. An attempted description in terms of the density matrix will consequently involve many undetermined parameters and will quickly become impractical for higher dimensions.

Instead, each set of measurement outcomes defines an equivalence class of compatible states and the specification of this equivalence class, say in the form of a matrix, holds all accessible information. The size of such a matrix description then depends on the number of measurements rather than the Hilbert-space dimensions, and can thus be substantially smaller than the density matrix. Moreover, this description enables the determination of entanglement conditions, by searching the equivalence class for separable states. To this end, we utilize the bipartite expectation value matrix (EVM) (cf.~\cite{rigas06a}) as an object which contains all measured information on the state.

In the following, we first present the general construction and properties of the EVM and then explain how to use it for a separability criterion.

\subsection{Construction of the Expectation Value Matrix}
Suppose Alice's and Bob's systems are associated with Hilbert spaces $\mathcal{H}_A$ and $\mathcal{H}_B$, respectively, and the joint system is in the state $\rho_{AB}$, acting on $\mathcal{H}_A \otimes \mathcal{H}_B$. Further suppose that Alice's and Bob's measurements are described by operators $\hat{A}_i^\dagger \hat A_k$ and $\hat{B}_j^\dagger \hat B_l$, respectively, for appropriate sets $\{\hat{A}_i\}$ and $\{ \hat{B}_j \}$. Then the entries of the bipartite EVM are defined as
\begin{equation}\label{eq:evm}
	[\chi(\rho_{AB})]_{ij,kl} = \text{Tr}(\rho_{AB} \hat{A}_i^\dagger \hat{A}_k \otimes \hat{B}_j^\dagger \hat{B}_l).
\end{equation}
A similar construction in terms of creation and annihilation operators leads to the so-called matrix of moments employed by Shchukin \emph{et al.}~to find criteria for non-classicality \cite{shchukin05b} and separability \cite{shchukin05a,miranowicz06b}. The latter method was extended to finite-dimensional systems in Ref.~\cite{miranowicz06suba}.\\

\emph{Observation 1:} The expectation value matrix defined by Eq.~(\ref{eq:evm}) is Hermitian and positive-semidefinite for all physical states $\rho_{AB}$.

\emph{Proof:}
The EVM is Hermitian since 
\begin{align*}
\chi^\dagger_{ij,kl} & = \chi^*_{kl,ij} \\
& = \text{Tr}(\rho_{AB} \hat{A}_k^\dagger \hat{A}_i \otimes \hat{B}_l^\dagger \hat{B}_j)^* \\
& = \text{Tr}(\rho_{AB} [\hat{A}_i^\dagger \hat{A}_k \otimes \hat{B}_j^\dagger \hat{B}_l]^\dagger)^* \\
& = \text{Tr}(\rho_{AB} \hat{A}_i^\dagger \hat{A}_k \otimes \hat{B}_j^\dagger \hat{B}_l) \\
& = \chi_{ij,kl}.
\end{align*}
Non-negativity is shown with the following argument: Given a physical state $\rho_{AB}$, the quantity Tr($ \rho_{AB} \hat{M}^\dagger \hat{M}$) is non-negative for all operators $\hat M$. Assuming the particular bipartite form $\hat M = \sum_{i,j} c_{ij} \hat{A}_i \otimes \hat{B}_j$ with the above sets $\{ \hat{A}_i \}$ and $\{ \hat{B}_i \}$, we find
\begin{align*}
0 & \le \text{Tr}(\rho_{AB} \hat{M}^\dagger \hat{M}) \\
& = \sum_{ij,kl} c_{ij}^* \text{Tr}(\rho_{AB} \hat{A}_i^\dagger \hat{A}_k \otimes \hat{B}_j^\dagger \hat{B}_l) c_{kl}\\
& = \sum_{ij,kl} c_{ij}^* [\chi(\rho_{AB})]_{ij,kl} \, c_{kl}, \\
& = \sum_{ij,kl} \vec{c}\ ^\dagger \chi(\rho_{AB}) \vec{c}, \quad \forall \vec{c}
\end{align*}
so $\chi(\rho_{AB})$ is a positive-semidefinite matrix. \hfill $\openbox$

Remarkably, the class of expectation value matrices contains, by suitable choices of the sets $\{ \hat{A}_i \}$ and $\{ \hat{B}_j \}$, the system's density matrix itself as well as the reduced density matrices. An EVM can also contain other useful quantities, such as the covariance matrix. An explicit construction of an EVM which contains $\rho_A$ and a variant of the covariance matrix of subsystem $B$ is used in Sec.~\ref{sec:apply}.\\

\subsection{Entanglement Criterion Based on Partial Transposition}
Quantum correlations exist in the measurement data if the equivalence class of states compatible with the observed EVM can be shown to contain no separable states. We will use the partial transposition to verify this. Due to the product structure of separable states \cite{werner89a}, transposition of either subsystem will result in another valid physical state. Therefore, if the partial transpose of a state $\sigma$ is not positive, $\sigma$ must be entangled \cite{peres96a}.

This property translates to the EVM. Since $\chi(\rho_{AB}) \ge 0$ for all physical states $\rho_{AB}$, violation of
\begin{equation}\label{eq:chipos}
\chi(\rho_{AB}^{T_A}) \ge 0
\end{equation}
is a sufficient condition for $\rho_{AB}$ to be entangled. The same holds for the transpose of subsystem $B$. In fact, a positivity requirement on a matrix, such as Eq.~(\ref{eq:chipos}), gives rise to a family of inequalities (cf.~\cite{shchukin05a}), since Eq.~(\ref{eq:chipos}) holds true if and only if all principle submatrices of $\chi(\rho_{AB}^{T_A})$ have non-negative determinants \cite{strang88a}. Negativity of any one of these sub-determinants provides a sufficient entanglement criterion, a non-trivial example of which will be given in Sec.~\ref{sec:evmstokes}.

To evaluate the EVM on $\rho_{AB}^{T_A}$, we use the invariance of the trace under partial transposition of its argument. The partial transpose of the state can thus be transferred to a transposition of Alice's measurement operators, i.e.:
\begin{equation}
	[\chi(\rho_{AB}^{T_A})]_{ij,kl} = \text{Tr}(\rho_{AB}[(\hat{A}_i^\dagger \hat{A}_k)^T \otimes \hat{B}_j^\dagger \hat{B}_l]).
\end{equation}
The usefulness of the entanglement criterion depends on the connection between the operators $\hat{A}_i^\dagger \hat{A}_k$ and their transpose. In the examples that we consider, it turns out that each transposed operator is merely a scalar multiple of one of the original operators, as illustrated in Sec.~\ref{sec:apply}.

Partial transposition is only one of many possible operations to detect entangled states. In fact, any positive but not completely positive map may be used as long as the adjoint map can act on the measurement operators in a meaningful way.

\subsection{Limited Information}
It may occur in experimental situations that not all entries of the EVM can be measured, however carefully the sets $\{ \hat{A}_i\}$ and $\{ \hat{B}_j \}$ are chosen. Entries which are not accessible must be represented by complex-valued free parameters. The criterion (Eq.~(\ref{eq:chipos})) can still be evaluated in this case by rephrasing it as follows: If it is not possible to find a set of free parameters such that
\begin{equation}\label{eq:cond}
	\chi(\rho_{AB}) \ge 0 \quad \text{and} \quad \chi(\rho_{AB}^{T_A}) \ge 0,
\end{equation}
$\rho_{AB}$ must have been entangled. This search through all possible parameter sets is efficiently done with semidefinite programming \cite{vandenberghe95a}. If it was known which parameter sets are compatible with physically valid states, the criterion (Eq.~(\ref{eq:cond})) with limited information would be as strong as the criterion (Eq.~(\ref{eq:chipos})) for the fully known EVM. This question for physically allowed parameter sets is, however, unanswered at present.\\

It is important to note that the verification scheme is, in principle, completely general, since neither the definition (Eq.~(\ref{eq:evm})) of the EVM nor the conditions in Eq.~(\ref{eq:cond}) contain assumptions about the measurements or the dimensionality.

\section{Application to a QKD Protocol with Binary Modulation}\label{sec:apply}
One of the strengths of using the EVM-based entanglement verification method is that high-dimensional states can be represented by a low-dimensional matrix, which will be of particular advantage in the continuous-variable regime.

We now apply the verification scheme to quantum key distribution. When effective entanglement can be verified in a particular implementation, the possibility of an intercept-resend attack is excluded.

The example we use to illustrate the entanglement verification method is a very simple protocol for quantum key distribution \cite{rigas06a,lorenz06a}. Alice generates a random classical bit-string which she wishes to share with Bob and use to distill secret key. To achieve this, she simply encodes each ``0" in a coherent state $\ket{\alpha}$ and each ``1" in $\ket{-\alpha}$, with $\alpha$ real and positive, and sends those signal states over a quantum channel to Bob. The notion of security arises form the non-orthogonality of the two signal states $\ket{\alpha}$ and $\ket{-\alpha}$.

In the equivalent entanglement based picture, Alice's source prepares bipartite states
\begin{equation}\label{eq:entbased}
	\ket{\psi}_{AB} = \frac{1}{\sqrt{2}} ( \ket{0}_A \ket{\alpha}_B + \ket{1}_A \ket{-\alpha}_B ).
\end{equation}
A subsequent projective measurement $\{ \ket{0}_A \bra{0}, \ket{1}_A \bra{1} \}$ effectively prepares the above signal states. Also, since system $A$ remains in the source, Alice has complete knowledge of the reduced density matrix $\rho_A$.

Under the action of the quantum channel the originally pure signal states evolve to mixed ones. We denote the resultant states at Bob's site by $\rho_0$ if Alice sent a 0 and by $\rho_1$ if Alice sent a 1. These conditional states are fed into a standard homodyne detection setup \cite{leonhardt97a} where they interfere with a strong coherent reference beam on a balanced beam splitter. The intensity difference of the output arms gives access to the field quadratures, defined as $\hat x = (\hat{a}^\dagger +\hat{a})/\sqrt{2}$ and $\hat p = \mathbbm{i}( \hat{a}^\dagger - \hat{a})/\sqrt{2}$. We assume here that for each incoming signal $\rho_i,\ i \in \{ 0,1 \}$, Bob can measure $\braket{\hat x}_i, \braket{\hat p}_i, \text{Var}_i(\hat x)$ and $\text{Var}_i(\hat p)$, where $\braket{\hat x}_i = \text{Tr}(\rho_i \hat{x})$ and $\text{Var}_i(\hat x) = \braket{\hat{x}^2}_i - \braket{\hat x}_i^2$ (and likewise for $\hat p$).

The existence of entanglement for the present quantum key distribution protocol has been investigated in \cite{rigas06a}. We will reproduce the results of this analysis using the method outlined above and show the equivalence of both approaches (see Appendix \ref{appsec:equiv}).

With the given measurements, the EVM is constructed in a straightforward way by choosing $\{ \hat{A}_i \} = \{ \ket{\psi}\bra{0}, \ket{\psi}\bra{1} \}$, where $\ket{\psi}$ is some generic (and irrelevant) qubit state, and $\{ \hat{B}_j \} = \{ \hat{\mathbbm{1}}_B, \hat x,\hat p \}$. Consequently, we obtain the $6 \times 6$ matrix
\[
 \chi(\rho_{AB}) = 
  \begin{bmatrix}
    \Big\langle |0\rangle\langle 0|\otimes B \Big\rangle_{\rho_{AB}} & 
   \Big\langle |0\rangle\langle 1|\otimes B \Big\rangle_{\rho_{AB}} \\
    \Big\langle |1\rangle\langle 0|\otimes B \Big\rangle_{\rho_{AB}} & 
 \Big\langle |1\rangle\langle 1|\otimes B \Big\rangle_{\rho_{AB}}
  \end{bmatrix}
\]
with
\[
B=\begin{bmatrix}
\hat{\mathbbm{1}}_B & \hat{x} & \hat{p}\\
\hat{x} & \hat{x}^2 & \hat{x}\hat{p}\\
\hat{p} & \hat{p}\hat{x} & \hat{p}^2
\end{bmatrix}.
\]
Note the special property that $\rho_A$ is included in the above EVM, so $\chi(\rho_{AB})$ indeed contains all information obtainable from the protocol.\\

We now evaluate the EVM on the partially transposed state $\rho_{AB}^{T_A}$. Taking an example entry and using the aforementioned invariance of the trace under partial transposition, we find
\begin{align}\nonumber
	[\chi (\rho_{AB}^{T_A}) ]_{2,6} & = \text{Tr}( \rho_{AB}^{T_A} [\ket{0} \bra{1} \otimes \hat{x}\hat{p}]) \\ \nonumber
	& = \text{Tr}( \rho_{AB} [\ket{0} \bra{1} \otimes \hat{x}\hat{p}]^{T_A}) \\ \nonumber
	& = \text{Tr}( \rho_{AB} [\ket{1} \bra{0} \otimes \hat{x}\hat{p}]) \\
	& = [\chi (\rho_{AB}) ]_{6,2}
\end{align}
and similarly for the other entries. Hence, $\chi(\rho_{AB}^{T_A})$ is the block transposition of $\chi(\rho_{AB})$.

Another important point is the undetermined entries of the EVM mentioned above. Both off-diagonal blocks, for example, are completely unknown with the exception of their (1,1) entries. Each unknown entry is represented by a complex number, so that the EVM is parameterized as
\begin{equation}\label{eq:chiopen}
	\chi = \frac{1}{2}
	\begin{bmatrix}
		\begin{array}{ccc}
			1 & \braket{\hat x}_0 & \braket{\hat p}_0 \\ 
			\braket{\hat x}_0 & \braket{\hat x^2}_0 & a+\frac{\mathbbm{i}}{2}\\
			\braket{\hat p}_0 & a-\frac{\mathbbm{i}}{2} & \braket{\hat p^2}_0
		\end{array} & 
		s \begin{bmatrix}
			1 & c & d \\
			c & f & g \\
			d & g-\mathbbm{i} & h
		\end{bmatrix} \\
		s \begin{bmatrix}
			1 & c^* & d^* \\
			c^* & f^* & g^*+\mathbbm{i} \\
			d^* & g^* & h^*
		\end{bmatrix} &
		\begin{array}{ccc}
			1 & \braket{\hat x}_1 & \braket{\hat p}_1 \\
			\braket{\hat x}_1 & \braket{\hat x^2}_1 & b+\frac{\mathbbm{i}}{2} \\
			\braket{\hat p}_1 & b-\frac{\mathbbm{i}}{2} & \braket{\hat p^2}_1
		\end{array}
	\end{bmatrix},
\end{equation}
where $s= \braket{-\alpha | \alpha}$, $a$ and $b$ are real parameters and $\{ c, \dots , h \}$ are complex. In the above parameterization, the commutator $[\hat{x},\hat{p}] = \mathbbm{i} \hat{\mathbbm{1}}$ was used to relate the products $\hat{x}\hat{p}$ and $\hat{p}\hat{x}$ to each other. Knowledge of the expectation value of the commutator is crucial for the success of the method. An example where this expectation value is not known is given in Section \ref{sec:evmstokes}.

Evaluating the entanglement criterion is now a matter of finding parameter sets $\{ a,b, \dots ,h \}$ such that the EVM satisfies the conditions given in Eq.~(\ref{eq:cond}). The task is simple in the sense that there exist efficient algorithms to solve this semidefinite feasibility problem \cite{lofberg04a,toh99a,sturm99a}.\\

Before presenting results of the numerical analysis, we comment on entangled states with positive partial transposition (PPT). Since our method is based on the negativity of the partial transposition, no such states can be detected. The following observation, however, shows that the given structure of signal states and measurements on system $A$ do not allow the detection of PPT entangled states on principle, independent of the method.

\emph{Observation 2:} (Compare \emph{Lemma 2} in \cite{rigas06a}) Let $\rho$ be a PPT entangled state not detected by the EVM method. Then there exists a separable state compatible with the same measured data.

\emph{Proof:} $\chi(\rho^{T_A})$ is the block transposition of $\chi(\rho)$ and therefore fits the same experimental data. Then so does the sum $(\chi(\rho^{T_A}) + \chi(\rho))/2$, which is the EVM of $\bar{\rho} = (\rho^{T_A} + \rho)/2$. This state $\bar{\rho}$ is invariant under transposition of subsystem $A$. Theorem $2$ in Ref.~\cite{kraus00a} states that any state of dimension $2 \times N$ which is invariant under transposition of subsystem $A$ is separable. We argue in Appendix \ref{appsec:2byN}	 that this theorem is applicable to the inifinite-dimensional state $\bar{\rho}$ and therefore, $\bar{\rho}$ is separable. \hfill \openbox

\subsection{Numerical Results for the Sufficient Condition}
Results of the numerical evaluation of the semidefinite program naturally rely on actually measured values of the known entries of the EVM. In the following, we simulate these experimental outcomes with a loss and noise model. The channel loss is modeled as a scaling factor $\sqrt{\eta}$ for the first quadrature moments, i.e., $\braket{\hat x}_{0/1}^{output} \rightarrow \sqrt{\eta} \braket{\hat x}_{0/1}^{input}$, and similarly for $\hat{p}$. Noise in the channel is assumed not to affect the first moments, but to act as a broadening in the observed variances. The variances are incorporated in the EVM (Eq.~(\ref{eq:chiopen})) by writing $\braket{\hat{x}^2}_{0/1} = \braket{\hat x}_{0/1}^2 + \text{Var}(\hat{x})$, and $\braket{ \hat{p}^2 }_{0/1} = \braket{\hat p}_{0/1}^2 + \text{Var}( \hat{p} )$. Unless otherwise stated, the noise is modeled to increase Var($\hat x$) and Var($\hat p$) equally (cf.~\cite{rigas06a}). A corresponding physical setup is a perfect channel with an inserted beam splitter, so that losses are attributed to the beam splitter's reflectivity. Feeding a thermal state into the second input port of the beam splitter can simulate the channel noise. Consequently, this noise model always leads to physical expectation values and actual experimental results are in good agreement with this model \cite{lorenz06a}. It should be noted that the verification method itself is independent of the choice of noise model.

Due to the particulars of the noise model, each distinct set of measurement outcomes is characterized, for a fixed loss value, by only two parameters. Namely, the input state overlap $\braket{-\alpha | \alpha}$, as chosen by Alice, and the quadrature variance measured by Bob. All parameter pairs with $0 \le \braket{-\alpha | \alpha} \le 1$ and $\text{Var}(\hat x) = \text{Var}(\hat p) \ge 1/2$ are physically valid.

\begin{figure}[ht]
	\centering
		\includegraphics[scale=0.27]{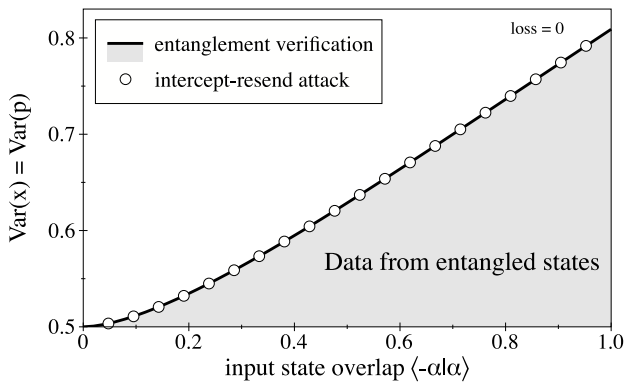}
	\caption{Input state overlap plotted against measured quadrature variances for a loss-less channel. Data points in the grey shaded area must stem from entangled states. The variance broadening induced by the optimal intercept-resend attack is indicated by the black dots.}
	\label{fig:quad}
\end{figure}

Figure \ref{fig:quad} shows which of these parameter pairs stem from entangled states. The plot is shown for $\eta = 1$. Corresponding curves for other values of the channel transmission have the same characteristic shape, but a smaller gradient, cf.~Fig.~2 in Ref.~\cite{rigas06a}. We see that entangled states are detected as long as the excess noise is not too high. The tolerable excess noise decreases for smaller input overlaps. Actual experiments can deliver values well within this region of entangled states \cite{lorenz06a}.\\

Note that an extension to heterodyne detection would allow the measurements of $\ket{i} \bra{i} \otimes \hat{x} \hat{p}$ and $\ket{i} \bra{i} \otimes \hat{p} \hat{x}$, for $i \in \{ 0,1 \}$. However, although this reduces the number of free parameters in the EVM, the results shown in Fig.~\ref{fig:quad} would remain the same for the given channel model.

\subsection{Refining the Method to be Necessary and Sufficient}
Whenever the conditions in Eq.~(\ref{eq:cond}) are met, the entanglement verification method is inconclusive, since the numerically-found solution may be unphysical. The question whether a given EVM corresponds to a physical state has no straightforward answer. Therefore, we supplement the method by searching for specific separable states compatible with those data sets satisfying Eq.~(\ref{eq:cond}). This search is equivalent to finding specific intercept-resend attacks \cite{curty04a}. In such an attack, an eavesdropper (Eve) measures out the signal states when they leave the source and resends appropriately chosen new states according to her measurement outcomes, thus acting as an entanglement-breaking channel \cite{horodecki03a}.

Intuitively, the non-orthogonality of the signal states protects against such attacks. When Eve tries to distinguish $\ket{\alpha}$ and $\ket{-\alpha}$ and reassign new signals accordingly, she inevitably makes errors which will be detectable in Bob's observations. In our case, these errors will manifest themselves in the form of increased uncertainties. Hence, for large measured variances we always expect to find an intercept-resend attack, in analogy to bit-error rates above 25\% in the qubit-based BB84 protocol \cite{moroder06a}.\\

A promising, and ultimately successful, choice of measurement for the eavesdropper is a minimum-error discrimination described in Ref.~\cite{helstrom76a}. For the signal states $\ket{\pm \alpha}$, the resulting error probability is given by
\begin{equation}
	e = \frac{1}{2} - \frac{1}{2} \sqrt{1-|\braket{-\alpha|\alpha}|^2}.
\end{equation}
Implementations of this minimum error discrimination measurement are discussed in Ref.~\cite{enk02c}.

In other words, if we assume Eve to resend either $\ket{\psi_0^E}$ or $\ket{\psi_1^E}$, she makes a wrong choice with probability $e$. As a result, Bob's states, conditioned on the original bit-values, become
\begin{align}
	\rho_0^{B,E} & = (1-e) \ket{\psi_0^E} \bra{\psi_0^E} + e \ket{\psi_1^E} \bra{\psi_1^E},\\
	 \rho_1^{B,E} & = (1-e) \ket{\psi_1^E} \bra{\psi_1^E} + e \ket{\psi_0^E} \bra{\psi_0^E}.
\end{align}
The exact form now depends on the choice of $\ket{\psi_{0/1}^E}$.

\begin{figure}[ht]
	\centering
		\includegraphics[scale=0.39]{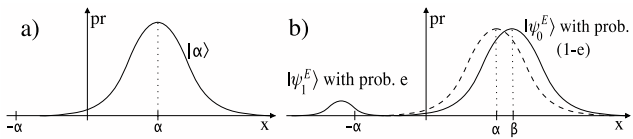}
	\caption{a) Marginal distribution for the input state $\ket{\psi_0} = \ket{\alpha}$. b) Bob's conditional state $\rho_0^{B,E}$ after the interaction of the eavesdropper. With probability $(1-e)$, Eve resent $\ket{\psi_0^E}$; with probability $e$, she erroneously resent $\ket{\psi_1^E}$.}
	\label{fig:evestates}
\end{figure}

We note here that since the original coherent amplitudes $\alpha$ have been chosen to be real, the expected variance broadening from the attack will only occur in the $\hat{x}$-direction (cf.~Fig.~\ref{fig:evestates}). However, Bob expects to observe equal variances in $\hat x$ and $\hat p$, which is achieved by the eavesdropper with appropriate squeezing operations. Hence, a natural choice of states is
\begin{equation}\label{eq:equad}
	\ket{\psi_{0/1}^E} =\hat{D}(\pm \beta) \hat{S}(r) \ket{0},
\end{equation}
where $\hat D$ and $\hat S$ are the displacement and squeezing operators, respectively, and $r$ is a real squeezing parameter \cite{barnett97b}.

For each input overlap, the parameters $\beta$ and $r$ can now be adjusted to reproduce the values $\braket{\hat x}_{0/1}$ and $\braket{\hat p}_{0/1}$ expected by Bob, while inducing the least possible excess variance. Denoting the input state overlap by $s$ and the channel transmission again by $\eta$, the analytic form of the resulting variances in Bob's measurements reads
\begin{align*}
	\text{Var}^{B,E}(\hat x) & = \text{Var}^{B,E}(\hat p) = \frac{1}{2} \big( f + \sqrt{f^2+1} \big),\\
	f & = \eta \frac{s^2 \ln s}{s^2 -1},
\end{align*}
and is shown in Fig.~\ref{fig:quad}. We see an exact coincidence with the entanglement verification curve. This coincidence holds for all channel loss values.

The combination of this attack with the entanglement verification method is necessary and sufficient to decide whether measured data is compatible with a separable state or not, for all possible data sets.

\subsection{Asymmetric Variances}
To assume equal variances for position and momentum in Bob's observations may be unjustified in some situations, but an extension of the above scheme to unbalanced excess noise is straightforward. The total measured variance will be comprised of input variance and excess noise, i.e.,
\[
\text{Var}^{tot}(\hat x) = \text{Var}^{in}(\hat x) + V^{ex}(\hat x) = \frac{1}{2} + V^{ex}(\hat x),
\]
and similarly for $\hat p$.
We find that, even in the case $V^{ex} (\hat x) \not = V^{ex}(\hat p)$, the maximum tolerable noise of entangled states and the minimum variance broadening induced by an attack (Eq.~(\ref{eq:equad})) coincide.

\subsection{Application to Quantum Memories}
We conclude this section with a short discussion on the application of the verification method to quantum memories. Recently, storage of continuous-variable states of light was realized \cite{appel07suba}. It was demonstrated that squeezing of vacuum input states was still observable in the output light.

If any two non-orthogonal input states $\ket{\phi_{0/1}}$ are chosen for testing a quantum memory, we can translate the storage process to an entanglement-based picture, in complete analogy with Eq.~(\ref{eq:entbased}), with $\ket{\pm \alpha}_B$ replaced by $\ket{\phi_{0/1}}$. Since the entanglement verification method only depends on the input-state overlap and the first and second moments of the quadratures of the output light, it is applicable to this situation, independent of the particular workings of the storage and readout process.

Verifiable entanglement in this picture is sufficient to ensure the quantum nature of the memory. Since this strategy requires no tomographically complete measurements, its experimental realization will require less resources.

\section{The EVM with Stokes Operators}\label{sec:evmstokes}

This section examines the protocol more closely by including the so-called local oscillator. This auxiliary light beam is used as a reference to define the signal states' phase and is necessary for homodyne detection. If the local oscillator is in a coherent state of high intensity compared to the signal, the detection setup can be used to measure the field quadratures, as in the previous section.

Ideally, the phase reference is prepared locally, but to ensure phase stability, it is common practice to split off the local oscillator in the signal preparation process and send it to Bob over the same quantum channel as the signal. The oscillator then inevitably passes through the insecure domain controlled by the eavesdropper and we cannot make unverified assumptions about its state at Bob's site. Consequently, the detection process should be regarded as a true two-mode measurement and we will employ the Stokes operators to describe it.

\subsection{Stokes Operators}
Typically, the Stokes operators are used to describe the quantum polarization of light using a decomposition into two modes of orthogonal polarization (see, e.g., Ref.~\cite{korolkova02a}). In our case, the modes are spatially separated and we will label them with subscripts ``$s$" and ``$LO$" to denote the signal and the local oscillator, respectively. In this language, the input signal states are denoted by
\begin{equation}\label{eq:sigstokes}
	\ket{\psi}_{0/1} = \ket{\alpha_{LO}, \pm \alpha_s}.
\end{equation}
The Stokes operators are defined in terms of mode operators as
\begin{align*}
\hat{S_1} & = \hat{a}_s^{\dagger} \hat{a}_s - \hat{a}_{LO}^\dagger \hat{a}_{LO}\\
\hat{S_2} & = \hat{a}_s^\dagger \hat{a}_{LO} + \hat{a}_{LO}^\dagger \hat{a}_s\\
\hat{S_3} & = i(\hat{a}_{LO}^\dagger \hat{a}_s - \hat{a}_s^\dagger \hat{a}_{LO}),
\end{align*}
and the total intensity is often included in the set:
\[
\hat{S_0} = \hat{a}_s^{\dagger} \hat{a}_s + \hat{a}_{LO}^\dagger \hat{a}_{LO}.
\]
Their commutation relations are those of the $\mathfrak{su}(2)$ algebra, namely,
\begin{equation}\label{eq:stcom}
	[\hat{S}_i, \hat{S}_j] = 2 \mathbbm{i} \epsilon_{ijk} \hat{S}_k
\end{equation}
for $\{ i,j,k\} = \{ 1,2,3 \}$, and $\epsilon_{ijk}$ is the Levi-Civita symbol. $\hat{S}_0$ commutes with the other Stokes operators. As a consequence of Eq.~(\ref{eq:stcom}), the minimum uncertainty product of any two Stokes operators depends on the expectation value of the third:
\begin{equation}
	\text{Var}(\hat{S}_i) \text{Var}(\hat{S}_j) \ge |\braket{\hat{S}_k}|^2.
\end{equation}
One other property worth mentioning is the operator identity
\begin{equation}
	\hat{S}_1^2 + \hat{S}_2^2 + \hat{S}_3^2 = \hat{S}_0^2 +2 \hat{S}_0.
\end{equation}

\subsection{Entanglement Verification}
The Stokes operators are measured by direct detection of intensity differences \cite{schnabel03a}. If the two incoming light modes are spatially separated, the setup is identical to that of a homodyne apparatus. Detecting the $\hat x$-quadrature then corresponds to an $\hat{S}_2$-measurement and the $\hat p$-quadrature to $\hat{S}_3$. We therefore assume that the two-mode detection process gives access to the expectation values and variances of $\hat{S}_2$ and $\hat{S}_3$.\\

An EVM is now constructed in complete analogy to the previous section, with $\{ \hat{B}_i \} = \{ \hat{\mathbbm{1}}, \hat{S}_2, \hat{S}_3 \}$, resulting in a matrix similar to (\ref{eq:chiopen}). The crucial difference is the commutator, which is unknown for this setting since it is proportional to the unmeasured quantity $\braket{\hat{S}_1}$. The known entries of this EVM are modeled to reflect actual experimental data. As before, losses can be modeled to downscale the first moments and noise is included as broadening of the variances of $\hat{S}_2$ and $\hat{S}_3$.
 
\begin{figure}[ht]
	\centering
		\includegraphics[scale=0.27]{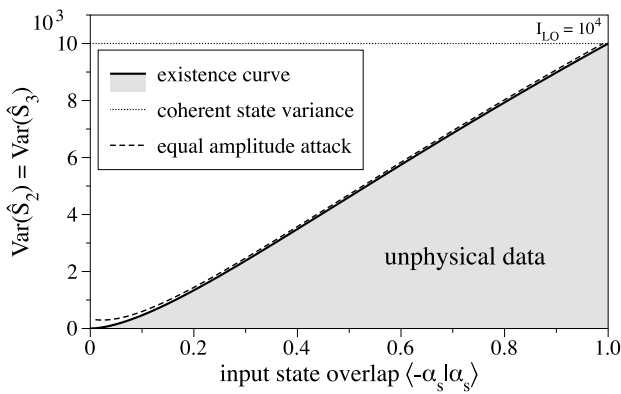}
	\caption{Signal state overlap plotted against the Stokes operator variance for a loss-less channel and local oscillator amplitude $\alpha_{LO}=100$. Only data points above the grey shaded area are physically allowed. Solid curve: entanglement verification without additional measurements. Dashed line: Intercept-resend attack with equal amplitudes in both modes. Dotted line: input state variance for reference.}
	\label{fig:stokes1}
\end{figure}

Numerical results are shown in Fig.~\ref{fig:stokes1}, again with the assumption that the observed variances in $\hat{S}_2$ and $\hat{S}_3$ are equal. Figure \ref{fig:stokes1} and all following figures are drawn for the loss-less case, since channel loss does not qualitatively change the results. An additional figure including channel loss can be found in Section \ref{sec:loss}.

There are two important points to note: Firstly, unlike the quadratures, the variances of the Stokes operators are not intensity-independent for coherent state input of the form (Eq.~(\ref{eq:sigstokes})). Rather, $\text{Var}(\hat{S}_i) = \braket{\hat{S}_0}$ holds true for all four Stokes operators when evaluated on coherent states \cite{korolkova02a}. Hence, the variances of the input states are also plotted. Due to the channel noise, actual measurement results are not expected below this coherent state curve.

Secondly, one can show that data sets below the entanglement verification curve are in fact unphysical. As mentioned above, every valid physical state must lead to a positive EVM, which implies that all principle minors of $\tilde \chi (\rho)$ must have non-negative determinants. Taking rows and columns (1,2,4) to form a particular submatrix, we require
\[
\det
\begin{pmatrix}
	1 & \braket{\hat{S}_2}_0 & s \\
	\braket{\hat{S}_2}_0 & \braket{\hat{S}_2^2}_0 & c \\
	s & c^* & 1
\end{pmatrix}
\ge 0,
\]
with $s= \braket{-\alpha | \alpha}$ as before and $c$ denotes a free parameter (cf. Eq.~(\ref{eq:chiopen})). Evaluating the determinant leads to
\begin{equation}\label{eq:exist1}
	\text{Var}_0 (\hat{S}_2) \ge \frac{\braket{\hat{S}_2}_0^2 s^2 - 2 s \braket{\hat{S}_2}_0 \text{Re}(c) + |c|^2}{1-s^2}.
\end{equation}
Positivity of the sub-determinant (1,4,5) translates to the similar condition
\begin{equation}\label{eq:exist2}
	\text{Var}_1 (\hat{S}_2) \ge \frac{\braket{\hat{S}_2}_1^2 s^2 - 2 s \braket{\hat{S}_2}_1 \text{Re}(c) + |c|^2}{1-s^2}.
\end{equation}
According to the noise model, $\braket{\hat{S_2}}_0 = -\braket{\hat{S}_2}_1$ and $\text{Var}_0(\hat{S}_2) = \text{Var}_1(\hat{S}_2) \equiv \text{Var}(\hat{S}_2)$. Adding (\ref{eq:exist1}) and (\ref{eq:exist2}) then gives the final condition
\begin{equation}\label{eq:exist}
	\text{Var}(\hat{S}_2) \ge \frac{s^2 \braket{\hat{S}_2}_0^2 + |c|^2}{1-s^2},
\end{equation}
which is obviously minimized for $c=0$. Plotting this minimum allowed variance against $s$ exactly coincides with the entanglement verification curve shown in Fig.~\ref{fig:stokes1}. Consequently, no entangled states are detected by our method in the current measurement setting.\\

To refine this last statement, we can again search for data pairs compatible with separable states by constructing a specific intercept-resend attack. The optimal measurement is the same as before, but the eavesdropper now resends
\begin{equation}\label{eq:evestokes}
	\ket{\psi^E}_{0/1} = \hat{D}_{LO}(\beta) \hat{D}_s(\pm \beta) \hat{S}_s (r) \ket{0}_{LO} \ket{0}_s,
\end{equation}
with appropriately chosen parameters. The intuition behind this particular choice is that equal coherent amplitudes for signal and local oscillator can lead to the same first moments of $\hat{S}_2$ and $\hat{S}_3$ as the original signals, but the total intensity is significantly decreased. This equal-amplitude attack can in principle lead to lower variances in Bob's observations than the input states would give, and this is indeed the case, as illustrated in Fig.~\ref{fig:stokes1}. In fact, in the limit of high photon numbers, the gap between existence curve and attack vanishes.\\

\subsection{Implications for Quantum Key Distribution}
Concluding this section, we find that entanglement verification is not possible with measurements of the first and second moments of $\hat{S}_2$ and $\hat{S}_3$ only. Almost all physically valid data sets admit an intercept-resend attack and are therefore compatible with separable states. No secure key can be distilled from such data.

Since homodyne detection is a standard tool for continuous-variable quantum communication, such as quantum key distribution with a Gaussian distribution of coherent signal states \cite{grosshans03a, heid07a}, the above attack may be a pitfall for a variety of protocols. Care must be taken that indeed the field quadratures are measured. Ways to ensure this are presented in the following section.

\section{Additional Measurements}\label{sec:addm}
We consider two additional experimentally feasible measurements and investigate their effect on the performance of entanglement detection.

\subsection{Measuring $\hat{S}_1$}
\begin{figure}[ht]
	\centering
		\includegraphics[scale=0.27]{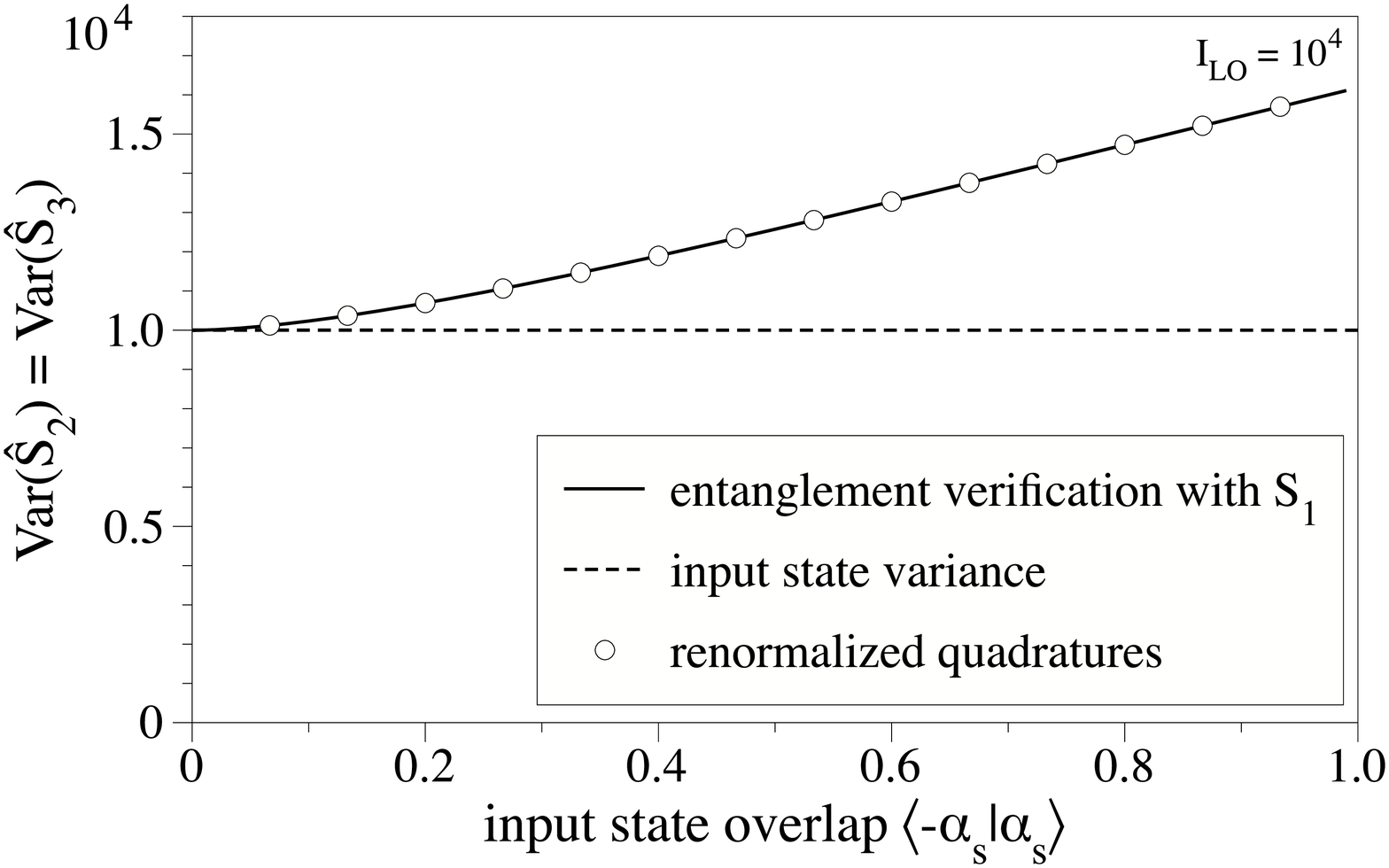}
	\caption{Signal state overlap plotted against the Stokes operator variance for the loss-less case with $\alpha_{LO}=100$. The renormalized quadrature variances (white circles) coincide with the Stokes operator variances when $\hat{S}_1$ is measured. Dashed line: input state variance for reference.}
	\label{fig:stokes2}
\end{figure}

A natural choice is the measurement of the third Stokes operator $\hat{S}_1$. Its expectation value and variance are easily included in the EVM by choosing the constructing set $\{ \hat{B}_i \} = \{\hat{\mathbbm{1}}, \hat{S}_1, \hat{S}_2, \hat{S}_3 \}$, enlarging $\chi(\rho_{AB})$ to an $8 \times 8$ matrix. It is now also possible to use the commutator (Eq.~(\ref{eq:stcom})) of $\hat{S}_2$ and $\hat{S}_3$ to relate products of the form $\hat{S}_2 \hat{S}_3$ and $\hat{S}_3 \hat{S}_2$ to each other. Since the entanglement verification method is otherwise identical to the preceding sections, we directly present the results in Fig.~\ref{fig:stokes2}.

There is a large area above the input state variance for which entanglement is ensured. In fact, if the results for the quadrature measurements are renormalized by the local oscillator intensity, they coincide with the present curve.

There is an intuitive explanation for these results. Since $\braket{\hat{S}_1}$ is the difference between the number of photons in the signal mode and that in the local oscillator mode, measuring $\braket{\hat{S}_1}$ provides a way of verifying that the phase reference is indeed much more intense than the signal. In this case, we expect to be able to measure the quadratures. A conceivably simpler way to achieve this verification is to monitor the local oscillator intensity separately.

However, these are truly additional measurements in the sense that the original setup must be modified.\\

Therefore, in Section \ref{sec:totint}, we briefly examine the situation where, instead of $\braket{\hat{S}_1}$, the total intensity $\braket{\hat{S}_0}$ is measured.

\subsection{Measuring the total intensity}\label{sec:totint}
The value of $\braket{\hat{S}_0}$ is obtained when the two photocurrents in $\hat{S}_2$ or $\hat{S}_3$ detection are added rather than subtracted. Depending on the experimental particulars, the total intensity may be available ``for free" with no change in the setup necessary \cite{wittmann07a}. The variance of $\hat{S}_0$ is not included in the analysis, since the intensity measurement is expected to be much more noisy than the measurements of the other Stokes operators.

Constructing an EVM from $\{ \hat{B}_i \} = \{\hat{\mathbbm{1}}, \hat{S}_0, \hat{S}_2, \hat{S}_3 \}$ gives no advantage over the case when $\hat{S}_0$ is not included. As far as entanglement detection is concerned, the results are identical to those shown in Fig.~\ref{fig:stokes1} above. However, the powerful equal-amplitude attack (Eq.~(\ref{eq:evestokes})) is not possible in this setting, and a new intercept-resend strategy must be constructed. The goal is to show a conceptual difference from quadrature measurements, i.e., to find an attack that explains data sets which would be attributed to entangled states, if quadrature detection was assumed. A corresponding attack can indeed be found. Its construction builds on the correspondence between a collection of photons in two possible polarization modes and an ensemble of spin-half particles aligned along some quantization axis. In fact, the Schwinger representation of spin operators \cite{schwinger65a} differs from the above definition of the Stokes operators only by factors of 2: $\hat{S}_{1,2,3} = 2 \hat{J}_{z,x,y}$.

For Eve's attack, we use the particular squeezing operation
\[
	\hat{U}(\mu) = \exp (-\frac{\mu}{8}(\hat{J}_+^2 - \hat{J}_-^2)),
\]
called two-axis countertwisting \cite{kitagawa93a}, where $\hat{J}_\pm = \hat{J}_x \pm \mathbbm{i} \hat{J}_y$. Her explicit states are given by
\begin{align}
	\rho_{0/1}^E & = p \ket{\psi_{0/1}^{(n^-)}} \bra{\psi_{0/1}^{(n^-)}} + (1-p) \ket{\psi_{0/1}^{(n^+)}} \bra{\psi_{0/1}^{(n^+)}} \\
	\ket{\psi_{0/1}^{(n^-)}} & = \hat{R}(\pm \frac{\pi}{2},0) \hat{U}(\mu) \ket{j,j} \\ \label{eq:nplus}
	\ket{\psi_{0/1}^{(n^+)}} & = \hat{R}(\pm \zeta,0) \hat{U}(\mu) \ket{j,j}
\end{align}
where $\ket{j,j}$ is an angular momentum eigenstate, $\zeta$ is a small rotation angle and $n^+$ ($n^-$) are photon numbers larger (smaller) than the total photon number Bob expects to measure. Both the rotation operator $\hat{R}(\theta,\phi)$ \cite{barnett97b} and the two-axis countertwisting operator conserve the total intensity.

\begin{figure}[ht]
	\centering
		\includegraphics[scale=0.27]{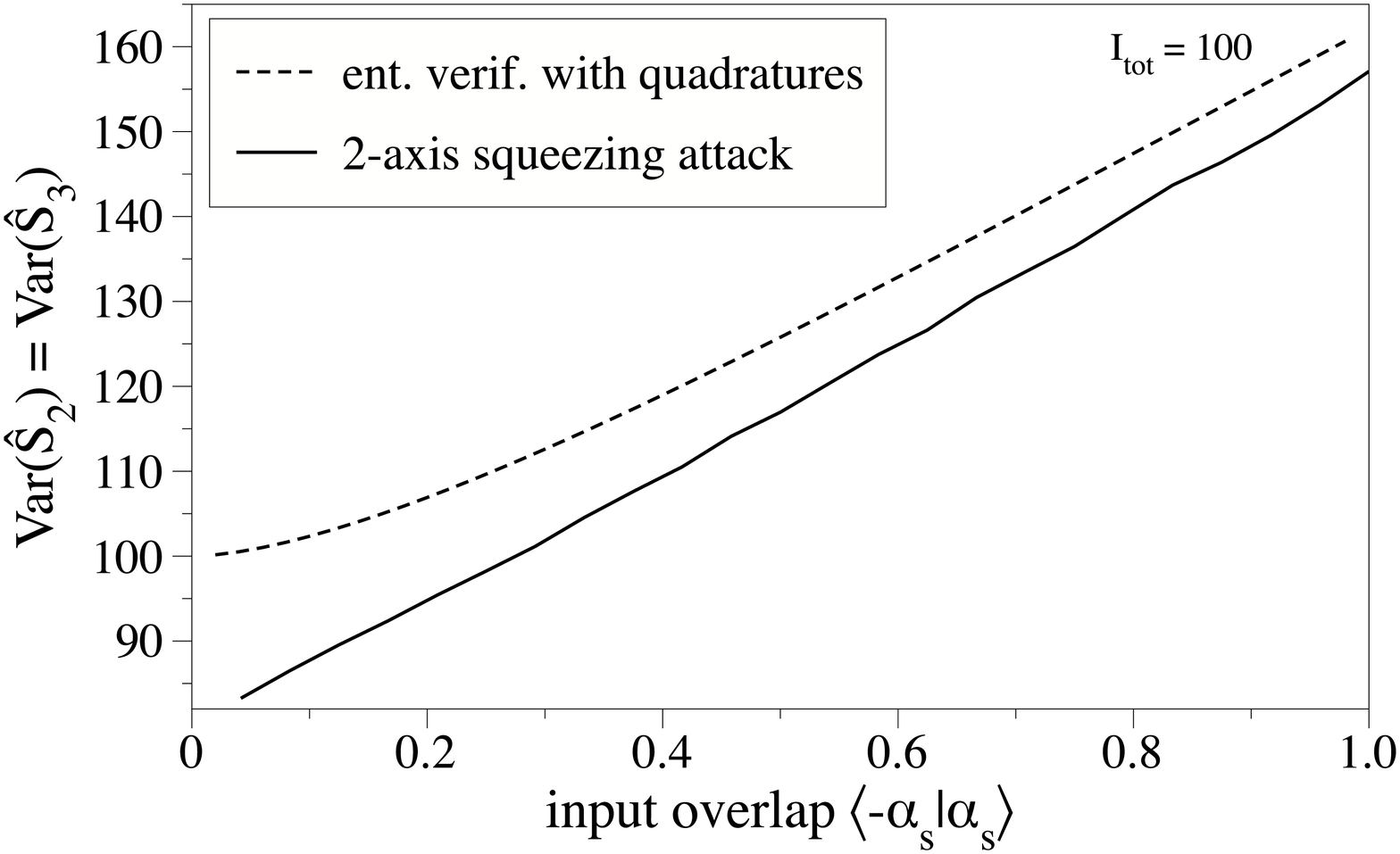}
	\caption{Performance of a 2-axis squeezing attack for 100 photons. Dashed line: Entanglement verification with quadrature measurements. Solid line: Intercept-resend attack with 2-axis twisting for Stokes operator detection and monitoring of the total intensity.}
	\label{fig:twoaxis}
\end{figure}

The minimum variance broadening induced by this attack is plotted in Fig.~\ref{fig:twoaxis} together with the entanglement verification curve for quadrature measurements. There are data sets from separable states well below the latter curve, so monitoring $\braket{\hat{S}_2}$, $\braket{\hat{S}_3}$, and the total intensity is not equivalent to quadrature detection and gives weaker results for entanglement verification. However, the attack was not evaluated for high photon numbers due to numerical limitations.

\begin{figure}[ht]
	\centering
		\includegraphics[scale=0.27]{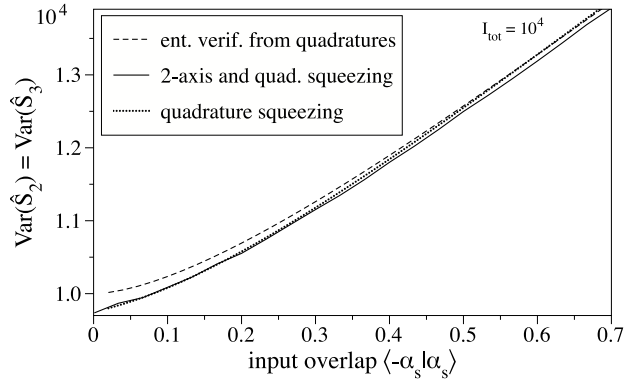}
	\caption{Different attacks for a total intensity of $10^4$. The dashed line results form entanglement verification with quadrature detection. Solid line: Attack with a mixture of 2-axis twisted and quadrature squeezed states for Stokes and total intensity measurements. Dotted line: The same attack with 2-mode squeezing instead of 2-axis countertwisting.}
	\label{fig:quadaxis}
\end{figure}

To be able to draw a comparison for more realistic intensities, we can replace Eq.~(\ref{eq:nplus}) by a quadrature squeezed state, i.e.
\[
\ket{\psi_{0/1}^{(n^+)}} = \hat{D}_{LO}(\tilde{\alpha}_{LO}) \hat{D}_s(\pm \tilde{\alpha}_s) \hat{S}_s(\tilde{r}^+) \ket{0}_{LO} \ket{0}_s.
\]
The need for numerical evaluation is completely removed by the further simplification
\[
\ket{\psi_{0/1}^{(n^-)}} = \hat{D}_{LO}(\tilde{\beta}) \hat{D}_s(\pm \tilde{\beta}) \hat{S}_{LO,s}(\tilde{r}^-) \ket{0}_{LO} \ket{0}_s,
\]
where $\hat{S}_{LO,s}$ is the two-mode squeezing operator. Although both of these simplifying steps deteriorate the effectiveness of the attack, they can still account for data points below the quadrature entanglement verification curve (Fig.~\ref{fig:quadaxis}). The resulting gap does increase with the total photon number, albeit at a much lower rate.\\

Still, a large number of data points is neither explained by the intercept-resend attack nor detected by the entanglement verification method. It remains unclear if the measurements do not provide enough information on the state or if the construction of the EVM is unsuited to verify entanglement for this particular scheme.

\section{Additional Figure for Channel Loss}\label{sec:loss}
\begin{figure}[h!]
	\centering
		\includegraphics[scale=0.27]{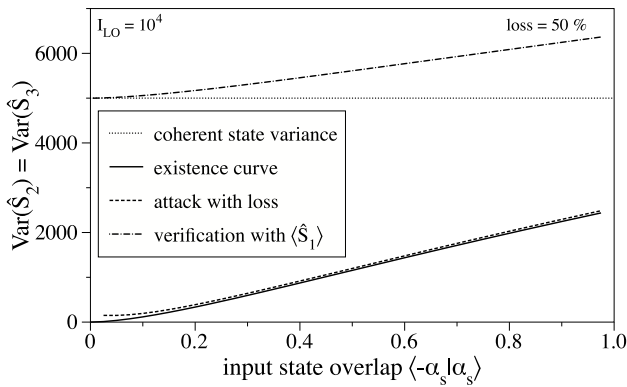}
	\caption{The above Stokes operator variances for a channel which transmits half of the input intensity.}
	\label{fig:losses}
\end{figure}

This final section concerns the effect of channel loss on all of the above results involving the Stokes operators. It is easily seen that the first moments of the Stokes operators scale as the channel transmission $\eta$, just as the reference coherent state variance. The existence curve, Eq.~(\ref{eq:exist}), however scales as $\eta^2$ and so does the corresponding intercept-resend attack. Typical results in the presence of loss are shown in Fig.~\ref{fig:losses} for 50 \% channel transmission, together with the entanglement verification including $\braket{\hat{S}_1}$.

\section{CONCLUSION}\label{sec:conclude}
In conclusion, we have constructed bipartite expectation value matrices as a description of systems with partial knowledge. Such matrices were used to formulate general sufficient criteria for entanglement, which allow the evaluation of partial transposition for, in principle, arbitrary dimensions and measurement operators.

The EVM method was applied to a communication protocol with coherent signal states and homodyne detection. With the assistance of an intercept-resend strategy, detection of the first and second moments of the field quadratures was sufficient to attribute all possible measurement outcomes to either separable or entangled states. A more complete description of the measurements in terms of Stokes operators revealed the inaptitude of the protocol to generate secret key. Two examples of additional measurements were considered to rectify this issue. Monitoring of the local oscillator intensity was found to recover the quadrature results. Instead, monitoring the total intensity showed data sets neither explainable by entanglement verification nor by the best known intercept-resend strategy. Thus, this last example demonstrated that the success of the presented method crucially depends on the choice and structure of Bob's measurements. This example also underlined the need for a criterion to decide which expectation value matrices are physically valid, which would result in stronger separability conditions. This question is, however, unanswered at present, and subject to future research.\\

Moreover, the presented method is proposed as a simple way to test quantum devices, which may be of particular interest for the continuous-variable storage of light. Two non-orthogonal signal states and detection of the first and second moments of the field quadratures of the output light suffice to verify the quantum operation of the memory, with no need for a complete reconstruction of the output density matrix. The particular light-atom interactions in the storage process are, as far as entanglement verification is concerned, not relevant.

\section{Acknowledgments} 
The authors would like to thank Geir-Ove Myhr, Matthias Heid, Volkher Scholz, and Johannes Rigas as well as Otfried G\"uhne and Natalia Korolkova for helpful discussions. The help on proving Observation 2 provided by Marco Piani and Maciej Lewenstein is gratefully acknowledged. The authors further thank Christoffer Wittmann and Dominique Elser for their help on experimental aspects. This work was funded by the European Union through the IST Integrated Project SECOQC and the IST-FET Integrated Project QAP. Further funding was received from the NSERC Innovation Platform Quantum Works and the NSERC Discovery Grant.

\appendix 
 
\section{Equivalent methods}\label{appsec:equiv}
In Ref.~\cite{rigas06a}, the constructed entanglement criteria are based on uncertainty relations. We will show that this approach is equivalent to our method based on partial transposition. Rigas \emph{et al.} \cite{rigas06a} constructed a symmetric version of the EVM defined as
\begin{equation}\label{eq:EVMdef}
  \chi^{s}(\rho) := 
  \begin{bmatrix}
    \Big\langle |0\rangle\langle 0|\otimes B^{s} \Big\rangle_\rho &
   \Big\langle |0\rangle\langle 1|\otimes B^{s} \Big\rangle_\rho \\
    \Big\langle |1\rangle\langle 0|\otimes B^{s} \Big\rangle_\rho &
 \Big\langle |1\rangle\langle 1|\otimes B^{s} \Big\rangle_\rho
  \end{bmatrix}
\end{equation}
with $B^{s}$ given by
\begin{equation}
  \label{eq:OpMatrixB}
  B^{s}:=
  \begin{bmatrix}
    \hat{\mathbbm{1}} & \hat x & \hat p\\
    \hat x & \hat x^2 & \frac{1}{2}(\hat x \hat p + \hat p \hat x)\\
   \hat p & \frac{1}{2}(\hat x \hat p + \hat p \hat x) & \hat p^2
  \end{bmatrix}.
\end{equation}
Then the inequalities
\begin{equation}\label{eq:Jcond}
	\chi^s (\rho) \pm \frac{\mathbbm{i}}{2} \rho_A \otimes \tilde{J} \ge 0,
\end{equation}
with
\[
\tilde{J} =
\begin{pmatrix}
0 & 0 & 0 \\
0 & 0 & -1 \\
0 & 1 & 0
\end{pmatrix},
\]
are satisfied by all separable states, and violation implies entanglement.\\

These two inequalities turn out to be equivalent to the conditions
\begin{equation}\label{eq:Hcond}
\chi (\rho) \ge 0 \qquad \text{and} \qquad \chi(\rho^{T_B}) \ge 0
\end{equation}
for the unsymmetrized EVM. Product entries can be split into their symmetrized product and commutator:
\begin{align}
\hat x \hat p & = \frac{1}{2}(\hat x \hat p + \hat p \hat x) + \frac{1}{2}[\hat x, \hat p]\\
& = \frac{1}{2}(\hat x \hat p + \hat p \hat x) + \frac{1}{2} \mathbbm{i},
\end{align}
so that $\chi(\rho) = \chi^s (\rho) - \mathbbm{i}/2 \ \rho_A \otimes \tilde{J}$.\\

Since Bob's operators are transposed according to the rule $\hat x \rightarrow \hat x$ and $\hat p \rightarrow -\hat p$, we find that
\begin{equation}
	\chi(\rho^{T_B}) = U \tilde \chi (\rho) U^\dagger,
\end{equation}
where $U = \mathbbm{1}_A \otimes \text{diag}[1,1,-1]$ and $\tilde \chi (\rho)$ is given by (\ref{eq:EVMdef}), with $B^s$ replaced by
\begin{equation}
  \tilde B:=
  \begin{bmatrix}
    \hat{\mathbbm{1}} & \hat x & \hat p\\
    \hat x & \hat x^2 & \hat p \hat x\\
   \hat p & \hat x \hat p & \hat p^2
  \end{bmatrix}.
\end{equation}
Then $\tilde \chi(\rho) = \chi^s (\rho) + \mathbbm{i}/2 \ \rho_A \otimes \tilde{J}$, i.e. the two conditions (\ref{eq:Jcond}) are, up to a local unitary transformation, equivalent to (\ref{eq:Hcond}).

\section{Supplementary Proof for Observation 2}\label{appsec:2byN}
We show that Theorem 2 in Ref.~\cite{kraus00a} applies to the infinite-dimensional case, i.e., to states $\rho_{AB}$ on $\mathcal{H}_A \otimes \mathcal{H}_B$ with $dim(\mathcal{H}_A) = 2$ and $dim(\mathcal{H}_B) = \infty$. That is, we prove the statement that if such a state $\rho_{AB}$ is invariant under transposition of subsystem $A$, it is separable. To this end, we approximate $\rho_{AB}$ by finite-dimensional states	
\begin{equation}
	\sigma_{AB}^{(N)} = \mathcal{N} (\mathbbm{1}_A \otimes P_B^{(N)}) \rho_{AB} (\mathbbm{1}_A \otimes P_B^{(N)}),
\end{equation}
with normalization factor $\mathcal{N}$ and the projection operator
\begin{equation}
	P_B^{(N)} = \sum_{k=1}^N \ket{k}_B \bra{k}.
\end{equation}
The vectors $\{ \ket{k}_B \}_{k=1}^\infty$ form a basis for system $B$. Intuitively,
\begin{align}
\| \rho_{AB} - \sigma_{AB}^{(N)} \|_{tr} & \equiv \delta_N, \\
\lim_{N \to \infty} \delta_N & = 0
\end{align}
where $\| \dots \|_{tr}$ denotes the trace norm. In other words, the sequence $\{ \delta_N \}$ is convergent. Therefore, $\forall \epsilon$, $\exists$ $n \in \mathbbm{N}$ such that $\rho_{AB}$ and $\sigma_{AB}^{(n)}$ are closer than $\epsilon$ in trace distance.

Furthermore,  $\sigma_{AB}^{(n)}$ is invariant under transposition of system $A$:
\begin{align}
	(\sigma_{AB}^{(n)} )^{T_A} & = \mathcal{N} [(\mathbbm{1}_A \otimes P_B^{(N)}) \rho_{AB} (\mathbbm{1}_A \otimes P_B^{(N)})]^{T_A} \\
	& = \mathcal{N} [(\mathbbm{1}_A \otimes P_B^{(N)})^{T_A} \rho_{AB}^{T_A} (\mathbbm{1}_A \otimes P_B^{(N)})^{T_A}] \\
	& = \mathcal{N} [(\mathbbm{1}_A \otimes P_B^{(N)}) \rho_{AB}^{T_A} (\mathbbm{1}_A \otimes P_B^{(N)})] \\
	& = \mathcal{N} [(\mathbbm{1}_A \otimes P_B^{(N)}) \rho_{AB} (\mathbbm{1}_A \otimes P_B^{(N)})] \\
	&= \sigma_{AB}^{(n)},
\end{align}
as by assumption, $\rho_{AB}^{T_A} = \rho_{AB}$. Therefore, $\sigma_{AB}^{(n)}$ is separable. Consequently, $\rho_{AB}$ is approximated by a convex combination of product states, so $\rho_{AB}$ is separable \cite{werner89a}. \hfill \openbox

\end{document}